\documentclass[10pt,letterpaper]{article}
\usepackage{opex3}
\usepackage{epstopdf}
\usepackage{amsmath}
\usepackage{cite}
\newcommand{\abs}[1]{\ensuremath{\left| #1 \right|}}

\newcommand{\be}[0]{\begin{equation}}
\newcommand{\ee}[0]{\end{equation}}
\newcommand{\bea}[0]{\begin{eqnarray}}
\newcommand{\eea}[0]{\end{eqnarray}}
\def\natp{ Nat.\ Phys. }
\def\njp{ New\ J.\ Phys. }

\begin{document}

\title{Influence of atmospheric turbulence on optical communications using orbital angular momentum for encoding}

\author{Mehul Malik,$^{1,*}$ Malcolm O'Sullivan,$^1$ Brandon Rodenburg,$^1$ Mohammad Mirhosseini,$^1$ Jonathan Leach,$^2$ Martin P. J. Lavery,$^3$ Miles J. Padgett,$^3$ and Robert W. Boyd$^{1,2}$}

\address{$^1$The Institute of Optics, University of Rochester,
Rochester, New York 14627 USA\\
$^2$Department of Physics, University of Ottawa, Ottawa, ON K1N 6N5 Canada\\
$^3$School of Physics and Astronomy, University of Glasgow, Glasgow, United Kingdom}

\email{*memalik@optics.rochester.edu} 



\begin{abstract}
We describe an experimental implementation of a free-space 11-dimensional communication system using orbital angular momentum (OAM) modes. This system has a maximum measured OAM channel capacity of 2.12 bits/photon. The effects of Kolmogorov thin-phase turbulence on the OAM channel capacity are quantified. We find that increasing the turbulence leads to a degradation of the channel capacity. We are able to mitigate the effects of turbulence by increasing the spacing between detected OAM modes. This study has implications for high-dimensional quantum key distribution (QKD) systems. We describe the sort of QKD system that could be built using our current technology.
\end{abstract}



\bibliographystyle{osajnl}


\section{Introduction}


Since Bennett and Brassard introduced the first quantum key distribution (QKD) protocol in 1984 \cite{Bennett:1984wv}, the field of QKD has rapidly developed to the extent that QKD systems are commercially available today. Secure transmission of a quantum key has been performed over 148.7 km of fiber \cite{Hiskett:2006kt} as well as over 144 km of free space \cite{Ursin:2007jj}. One of the limiting aspects of these key distribution systems is that they use the polarization degree of freedom of the photon to encode information \cite{Hiskett:2006kt,Ursin:2007jj}. The use of polarization encoding limits the maximum amount of information that can be encoded on each photon to one bit. In addition, it places a low bound on the amount of error an eavesdropper can introduce without compromising the security of the transmission \cite{Bourennane:2002uo}. Because of these limitations, there has been great interest in exploring other ways to encode information on a photon that would allow for higher data transmission rates and increased security \cite{Groeblacher:2005ec,Cerf:2002fp}.

Here we report results on the use of orbital angular momentum (OAM) modes of a photon for optical communication. The motivation for doing so is that OAM modes span an infinite-dimensional basis. Hence, there is no limit to how much information one can send per photon in such a system. The large dimensionality of this protocol also provides a much higher level of security than the two-state approach \cite{Bourennane:2002uo}. However, in a practical communication system using OAM modes, the maximum number of modes that can be used is limited by the size of the limiting aperture in the system. This occurs because the radius of an OAM mode increases with the mode number. In addition, since these are spatial modes, they are highly susceptible to turbulence. Recently, there have been several theoretical studies on how atmospheric turbulence affects OAM modes \cite{Paterson:2005wj,Tyler:2009wz,Smith:2006ek,Gbur:2008fb,Roux:2011dr}. In this paper, we experimentally study the effects of atmospheric turbulence on the channel capacity of an OAM communication channel at high light levels. Also, we describe a high-dimensional QKD system that could be built using OAM modes for encoding.

\section{Quantum key distribution system}

In a conventional QKD system, the sender (Alice) and receiver (Bob) must have at least two mutually unbiased bases (MUBs) in which to send and receive their key. The maximum number of MUBs in an $N$-dimensional QKD system is equal to $N+1$, for when $N$ is a prime number \cite{Wootters:1989ba}. For the polarization-based implementation of the BB84 protocol \cite{Bennett:1984wv}, $N$ is equal to 2 and there are three available MUBs. Here we describe a proposed OAM-based QKD system with a dimensionality $d=11$, and 12 possible mutually unbiased bases. Out of these, we choose to use only two in order to maximize our key transmission rate, which is inversely proportional to the number of MUBs that are used. The first basis is made up of 11 OAM modes and is called the OAM basis. The OAM modes have the form

\be\Psi_{\textrm{\tiny OAM}}^{l,0}=A_0W(r/R)\exp{(il\theta)}\ee

\begin{figure}[t]
\centering\includegraphics[scale = .66]{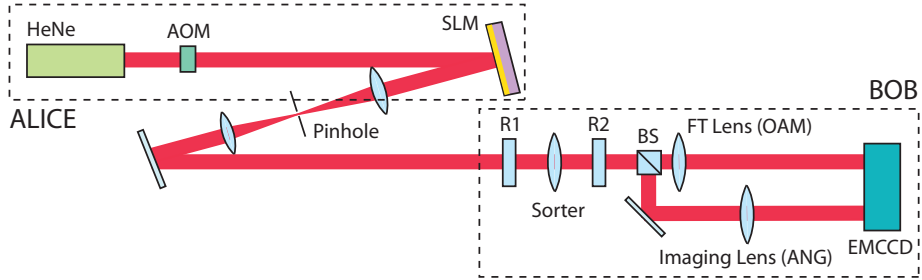}
\caption{Proposed QKD Setup. Our source is a HeNe laser operating at 633 nm that is modulated by an acousto-optic modulator (AOM) to carve out single-photon pulses. The OAM and ANG states, as well as the turbulence phase screens are generated by the spatial light modulator (SLM) and $4f$ system. R1 and R2 are custom refractive elements used to sort the OAM and ANG modes, and the beam splitter (BS) acts as the passive basis selector. The modes are detected on an electron-multiplying camera (EMCCD).}\label{setup}
\end{figure}

\noindent where $A_0$ is the spatially uniform field amplitude, $W(x)$ is an aperture function such that $W(x)=1$ for $\abs{x}\leq 1$ and zero otherwise, $r$ and $\theta$ are the radial and azimuthal coordinates, and $l$ is the OAM quantum number. The second basis is called the ANG basis and is made up of angular modes, which are a linear combination of the 11 OAM modes and have the form

\be \Psi_{\textrm{\tiny ANG}}^n=\frac{1}{\sqrt{11}}\sum_{l=-5}^5\Psi_{\textrm{\tiny OAM}}^{l,0}\exp{(i2\pi nl/11)}.\ee

These modes are so named because their intensity profile looks like an angular slice that moves around the center of the beam as one changes the relative phases of the component OAM modes. In order to characterize the effects of turbulence on our system, we choose to use pure vortex modes with a circular aperture function instead of Laguerre-Gauss modes in our system. We do this because our theoretical approach closely follows that of reference \cite{Tyler:2009wz}, which is also based on pure vortex modes.

A schematic of our proposed QKD system is shown in Fig. 1. We use a HeNe laser modulated by an acousto-optical modulator (AOM) as our source. The AOM carves out single-photon pulses for the QKD protocol. The OAM and ANG modes are generated by Alice using a hologram generated on a Holoeye PLUTO phase-only spatial light modulator (SLM) in conjunction with a $4f$ system \cite{Arrizon:2007wl, Gruneisen:2008bu}. Figure 2 shows some of the holograms we used and the OAM and ANG modes generated by them.

The generated modes are then imaged across a free-space channel onto Bob's detection plane. One of the key hurdles to using OAM modes to perform QKD has been the need of a method of efficiently sorting single photons carrying OAM modes. This problem was recently overcome by a method that uses two phase screens to conformally map an input OAM state to a shifted spot in the fourier plane of a lens \cite{Berkhout:2010cb}. In our setup, we are using custom refractive elements designed according to this method in order to carry out efficient sorting of OAM modes \cite{Lavery:2012hi}. The OAM sorter maps the input OAM modes to a row of spots on our detector. 

\begin{figure}[t]
\centering\includegraphics[scale = .65]{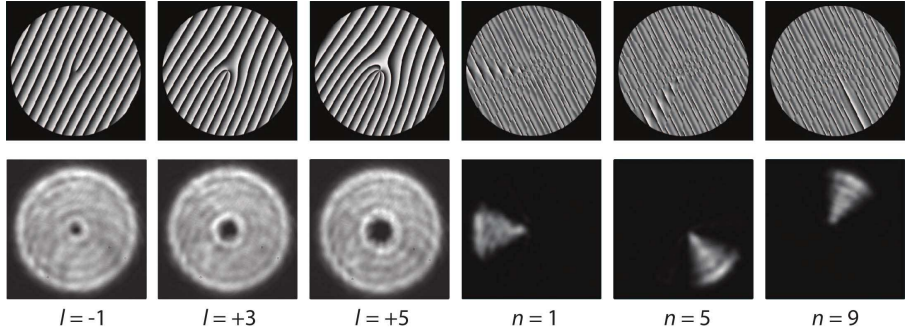}
\caption{Upper row: Holograms written onto the SLM to generate OAM and ANG modes. Lower row: CCD images of the corresponding modes generated (OAM mode number $l=-1$, 3, and 5, and ANG mode number $n=1$, 5, and 9).}\label{modes}
\end{figure}


Since the complex ANG modes are mostly non-overlapping in intensity, they can be sorted based on their angular position. For convenience, we use the first element (R1) of the OAM sorter to map the angular intensities to a partially overlapping row of spots, which are imaged onto the detector. The beam splitter (BS) acts a passive selector of Bob's measurement basis, randomly measuring photons in either the OAM or ANG basis. We use an electron-multiplying CCD (EMCCD) as our single-photon detector. While we intend on using this setup to perform QKD in the future, currently we are using it at high light levels for the purposes of characterizing the effects of turbulence on the OAM channel.

\section{Effects of turbulence on channel capacity}

Atmospheric turbulence plays a large role in determining how practical a free-space communication system may be. Here, we study the effects of weak turbulence on the OAM communication channel. In our system, the aperture of the sender is imaged onto the aperture of the receiver. For such a system, turbulence can be modeled as a thin-phase screen in the sending aperture \cite{Young:1974se}. The effects of turbulence are characterized in terms of crosstalk in the OAM and ANG channels. However, the effects of thin phase turbulence on the ANG channel cannot be studied in our system because of two reasons. First, the sorter is limited in that it only maps the non-overlapping intensities of the complex ANG modes to a row of spots. Any phase information in the ANG modes is not conveyed. Second, the thin-phase turbulence is in an image plane of the ANG channel. Since the ANG modes are imaged from the SLM to the sorter element R1, and then intensity-mapped to the detector, the effects of turbulence added at the SLM are not apparent in the sorted ANG modes.




However, we can study the effects of turbulence on the OAM communication channel, as the sorter sorts these modes based on their orthogonal phase distributions. The weak turbulence regime can be approximated by the Kolmogorov thin-phase model of turbulence. The phase structure function in this model obeys the relationship 

\be\langle[\phi(\mathbf{r_1})-\phi(\mathbf{r_2})]^2\rangle=6.88\abs{\frac{\mathbf{r_1-r_2}}{r_0}}^{5/3}\ee

\noindent where $r_0$ is Fried's coherence diameter \cite{Fried:1965gu}, a measure of the transverse distance scale over which refractive index correlations remain correlated. We generate turbulence phase screens based on this model \cite{Harding:1999hv} and add them to the pattern written on the SLM. An example of three Kolmogorov phase screens with different values of $r_0$ is shown in Fig. 3. The crosstalk introduced by this type of turbulence into an OAM channel can be written in the form \cite{Tyler:2009wz}

\begin{figure}[t]
\centering\includegraphics[scale = .55]{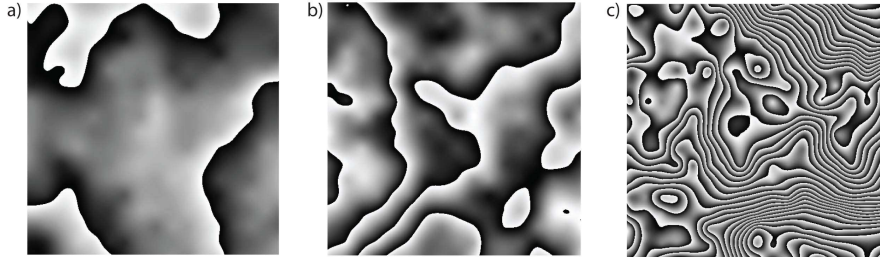}
\caption{Sample Kolmogorov turbulence phase screens that are added onto the SLM for three different values of turbulence strength, $D/r_0=$(a) 5.12, (b) 10.25, and (c) 102.5. Each screen has a resolution of 512$\times$512 pixels and the phase wraps from 0 to 2$\pi$.}\label{turb}
\end{figure}

\be P_{ds} = \frac{1}{\pi}\int_0^1\rho\mathrm{d}\rho\int_0^{2\pi}\mathrm{d}\theta e^{-3.44(D/r_0)^{5/3}(\rho\sin{\theta/2})^{5/3}}\cos{[(d-s)\theta]}\ee
\begin{figure}[b!]
\centering\includegraphics[scale = .5]{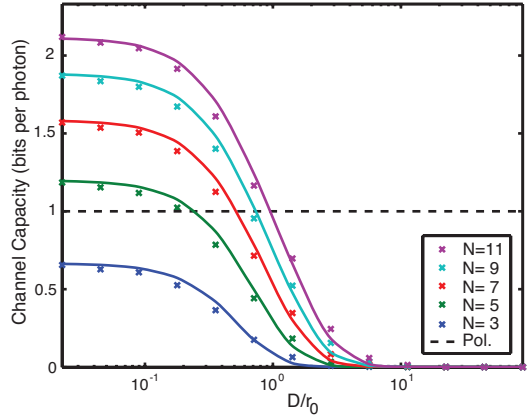}
\caption{Influence of Kolmogorov turbulence on the channel capacity of an OAM communication channel for a system dimensionality of $N=3,5,7,9$ and $11$. The channel capacity of a system using polarization modes is plotted as a dotted line for comparison.}\label{graph1}
\end{figure}

\noindent where $\rho=r/R$, and $P_{ds}$ is the conditional probability that Bob detects an OAM mode with quantum number $d$ given that Alice sent an OAM mode with quantum number $s$. Here, D is the diameter of the receiving aperture. From these probabilities, we can calculate the channel capacity of our OAM channel as
\bea C&=&\textrm{max}[H(x)-H(x|y)]\nonumber\\
&=&\underset{\{\mathit{P_s}\}}{\mathrm{max}}\bigg[-\sum_s P_s\log_2{(P_s)}+\sum_s P_s \sum_d P_{ds}\log_2(P_{ds})\bigg].
\eea

\noindent Here, $N$ is the total number of modes in our system and $P_s$ is the probability that Alice sent mode $s$. Experimental data for the channel capacity of the OAM communication channel with $N=11, 9, 7, 5, \textrm{ and }3$ are shown in Fig. 4. The channel capacity, $C$, is plotted as a function of the turbulence strength, $D/r_0$, where $D$ is the diameter of the photon field at the SLM and $r_0$ is Fried's coherence diameter. For each data point, we average over 20 different turbulence screens. The theoretical crosstalk is calculated by eq. (4) and corrected for the inherent crosstalk introduced by the sorter \cite{Berkhout:2010cb}. The theoretical channel capacity as calculated by eq. (5) is plotted in Fig. 4, as well as the channel capacity of a polarization-based two-dimensional system, which is essentially immune to the effects of turbulence. 

As can be seen, the maximum channel capacity for $N=11$ is much lower than the theoretical maximum of $\log_2(11)=3.46$ bits/photon. This is because the OAM sorter is not perfect and introduces some inherent crosstalk \cite{Berkhout:2010cb}. Nonetheless, our higher dimensional systems perform much better than the polarization-based system below a certain turbulence strength. This is indicated in Fig. 4 by the point at which the channel capacity curve for an $N$-dimensional system crosses the dotted line for a polarization-based system. At a turbulence strength of $D/r_0=10$, the channel capacity vanishes for all systems.


\begin{figure}[t!]
\centering\includegraphics[scale = .5]{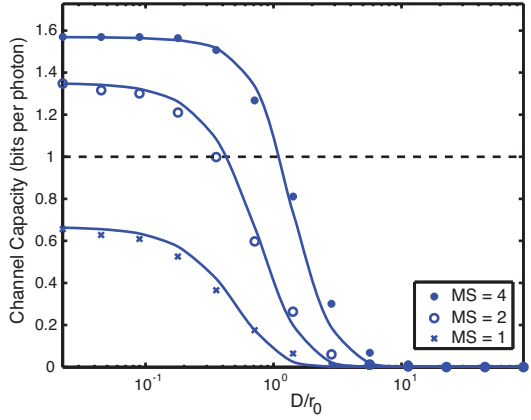}
\caption{Effects of changing the spacing between detected modes (MS) on the channel capacity of a 3-dimensional OAM communication system in the presence of atmospheric turbulence. For a mode-spacing of 4, the maximum channel capacity of a 3-dimensional system is seen to approach the theoretical maximum of $\log_2(3)=1.585$.}\label{graph1}
\end{figure}

Since OAM modes constitute an infinite-dimensional basis, one can reduce the effects of the inherent sorter crosstalk and turbulence by increasing the spacing (MS) between detected modes to greater than one. We study this effect on a subset of our system with N = 3, for three different mode-spacings (MS = 1, 2 and 4). It is clear from Fig. 5 that increasing the mode spacing between detected modes boosts the initial channel capacity and makes our system slightly more resistant to turbulence. The maximum channel capacity for a mode-spacing of four is close to the theoretical maximum of $\log_2(3)=1.585$. In addition, the channel capacity for this mode-spacing starts decreasing at a value of $D/r_0$ that is almost an order of magnitude greater than that for a mode-spacing of one.

\section{Conclusions}

We characterized the effects of Kolmogorov thin-phase turbulence on the channel capacity of an OAM-based communication system at high light levels. We found that turbulence affects the channel capacity drastically, and the effects can be mitigated by increasing the mode-spacing. We were not able to study the effects of thin-phase turbulence on the ANG channel due to the inherent limitations of our mode sorter. We are currently working on improving the performance of our mode sorter and extending our study to thick-phase turbulence. We would like to thank Dr. D. J. Gauthier for useful discussions. This work was supported by the DARPA/DSO InPho program and the Canadian Excellence Research Chair (CERC) program.

\end{document}